\documentstyle[amssymb,preprint,aps]{revtex}
\tightenlines
\begin{document}
\title{Schemes for Parallel Quantum Computation \\
Without Local Control of Qubits}
\author{S. C. Benjamin, Univ. of Oxford.}
\author{s.benjamin@physics.ox.ac.uk}
\date{\today}
\maketitle

\begin{abstract}
Typical quantum computing schemes require transformations (`gates') to be
targeted at specific elements (`qubits'). In many physical systems, direct
targeting is difficult to achieve; an alternative is to encode local gates
into globally applied transformations. Here we demonstrate the minimum
physical requirements for such an approach: a one-dimensional array composed
of two alternating `types' of two-state system. Each system need be
sensitive only to the {\em net} state of its nearest neighbors, i.e. the
number in state `$\uparrow $' minus the number in `$\downarrow $'.
Additionally, we show that all such arrays can perform quite general {\em %
parallel} operations. A broad range of physical systems and interactions are
suitable: we highlight two examples.

\newpage
\end{abstract}

Presently there is tremendous interest in the new field of quantum
computation. Information is recognized as a physical quantity, with its
representation and processing being governed by the laws of quantum
mechanics.\ Rather than `bits', the fundamental units of classical
information theory, we instead employ `qubits' which represent a general
quantum superposition of `0' and `1'. A computation on a device containing $%
N $ qubits is a sequence of unitary transformations within its $2^{N}$
dimensional Hilbert space. Researchers have already discovered quantum
algorithms which exploit state superposition, entanglement and interference
in order to solve certain problems more quickly than any known classical
procedure \cite{SteaneRev}. Two important cases are those of factoring large
numbers, where the quantum device has an exponential speed advantage \cite
{factoring}, and the task of searching among $N$ elements, where a classical
device requires time of order $N$ but the quantum device requires only $%
\sqrt{N}$ \cite{search1}, or $\sqrt[3]{N}$ with a corrresponding size cost 
\cite{our2DCA}.

Efforts toward experimental realization of a quantum computer (QC) have
focused principally on NMR and atomic trap implementations \cite
{NMR,SteaneIon}. Numerous recent proposals have also drawn attention to
possible solid state realizations \cite{Kane,DiVen,our2DCA}. Typically such
proposals demand manipulation of the Hamiltonian locally, on the scale of
the individual component qubits. However this is not a fundamental
requirement: it can be sufficient to apply only global manipulations to
which all elements are subjected simultaneously. This would be a highly
desirable economy for many implementations, because it would lower the
number of channels by which the computer interacts with its environment, and
hence reduce the decoherence rate. Moreover, it may enable new
implementations where it is difficult or impossible to perform individual
addressing (e.g. quantum dot arrays may be driven by EM radiation of a
wavelength far greater than the dot-dot separation). Lloyd has suggested one
such model \cite{LloydScience,sethLong} based on a one-dimensional cellular
automaton (CA). The model consists of a line of `cells',\ where each cell is
a quantum system possessing two long-lived internal eigenstates. The
algorithm is represented by a series of update `rules' which are applied
globally\ to all cells, so that there is {\em no need} to address units
individually. To realize LLoyd's CA model one would need to produce three
`types' of cell in the pattern $ABCABC$..., and moreover one must find a
means of applying asymmetric rules such as, ``all cells of type $A$ now
invert their state if, and only if, the left neighbor is in state 1 and the
right is in state 0''. Clearly it is important to know if these are the
minimum physical requirements. The present work demonstrates that they can
in fact be relaxed significantly, to {\em two} cell types {\em without} the
ability to distinguish the left neighbor from the right. These are the
minimum requirements for any globally-driven system, given that we must have
more than one cell `type' \cite{need2Types}. The simplifications enhance the
practicality of the model; neighbor indistinguishability is particularly
significant in broadening the range of potential implementations, two of
which we later discuss. This paper also provides a mechanism, compatible
with any CA\ computer, for performing operations in parallel. We note the
implications in terms of device size and speed. Parallelism may be essential
for quantum error correction schemes to function efficiently \cite
{parErrorCor}.

Our scheme consists of a two `types' of cell, A and B, alternating along a
one-dimensional array. Each cell has two internal eigenstates $|\downarrow
\rangle $ and $|\uparrow \rangle $, and can represent any quantum
superposition of these states. Each qubit of quantum information is
represented by {\em four} consecutive cells: the qubit basis state $%
|0\rangle $ is represented by $|\uparrow \uparrow \downarrow \downarrow
\rangle $ whilst the state $|1\rangle $ is represented by $|\downarrow
\downarrow \uparrow \uparrow \rangle $. The basis states of a qubit $X$ can
therefore be compactly written as $\overline{xx}xx$, where $x$ corresponds
to $\downarrow $\ if $X=0$ and $\uparrow $\ if $X=1$, with the opposite
applying to $\overline{x}$. Figure 1(a) shows an array containing three
qubits, each pair being separated by spacer cells in the $|\downarrow
\rangle $ state (the minimum acceptable spacing is four cells \cite{webFigs}%
, but eight are used here for clarity). The array is subject to update rules
specified by the notation $A_{f}^{{\bf U}}$ which means, each cell of type $%
A $ is subjected to unitary transform ${\bf U}$ if, and only if, its `field'
has value $f$. When the ${\bf U}$ is omitted a simple inversion is implied, $%
|\downarrow \rangle $ $\rightleftarrows $ $|\uparrow \rangle $. The `field'
is defined as the number of nearest neighbors in state $|\uparrow \rangle $\
minus the number in state $|\downarrow \rangle $. This is the proper control
variable since in a physical realization the cells will be aware of their
neighbors through the {\em net} effect of, for example, their electrostatic
fields.

In classical computing we have the idea of a universal set of gates, i.e. a
set of elementary operations (such as AND, OR, NOT) which are sufficient to
represent any classical algorithm. In quantum computing the same concept
applies. We first consider the general `one-qubit gate', i.e. any chosen
unitary transform applied to any particular qubit regardless of the other
qubits in the computer. How can we single out one qubit, given that the
array is structurally regular and our rules must be sent to all elements
globally? One solution \cite{LloydScience} is to introduce a `control unit'
(CU). Our CU is represented by six consecutive cells in the pattern $%
\uparrow \uparrow \downarrow \downarrow \uparrow \uparrow $, which exists
only in one place along the array. In Figs. 1(b) and 1(c) we utilize the CU\
by applying updates which move it relative to the qubits, together with an
update sequence which has a net effect only on the qubit nearest the CU.
Clearly, by varying the update sequence we could have transformed another
qubit, or indeed simply moved the CU without altering any of them. Thus we
can implement a general one-qubit gate. For a universal set we require a
two-qubit gate: the `control-${\bf U}$' is more than sufficient. This gate
applies the transform ${\bf U}$ to a certain qubit (referred to as the
`target') if, and only if, a second qubit (the `control') is in state 1.
Figure 2 shows the implementation of this gate schematically; an explicit
depiction analogous to Fig.1(b) is also available \cite{webFigs}. The `cost'
of restricting ourselves to global manipulations is now apparent: each qubit
requires a total of eight physical cells (four for the encoding plus four
spacers), and a 1-qubit gate requires about ten elementary pulses. These
numbers would be somewhat smaller if we permitted ourselves more cell types
and/or more complex interactions \cite{LloydScience}.

To input information we may exploit the cells at the ends of the array \cite
{LloydScience}, for whom the possible values of the `field' variable are $%
\{1,-1\}$ (in contrast to the $\{-2,0,2\}$ values for all other cells). We
can use the updates $\{$ $A_{-1}$,$A_{1}$,$B_{-1}$,$B_{1}\}$ to manipulate
the end states, and the other updates to shift-load those states into the
array. The means of output will depend on the available measurement
techniques. If a cell on one end were associated with a measuring device,
then one would first swap the qubit to be measured with the qubit nearest
the end (by a series of three CNOTs, for example), then move it onto the end
cell by the reverse of the input technique. A superior output procedure
would be possible if, for example, the cells of type $B$ had some third
state `$\rightsquigarrow $' exhibiting rapid spontaneous decay to the $%
\downarrow $ state. Then we could measure the state of a qubit anywhere
along the array using the 1-qubit gate of Fig. 1(b) and choosing

${\bf U}$=$\left( 
\begin{array}{ccc}
0 & 0 & 1 \\ 
0 & 1 & 0 \\ 
1 & 0 & 0
\end{array}
\right) $ in the basis $\{\downarrow ,\uparrow ,\rightsquigarrow \}$.

\noindent If the subject qubit was previously in state 1, then the
transformation would leave its representative cell in the unstable state $%
\rightsquigarrow $. From there it would decay back to $\downarrow $\ with an
emission. The presence (or absence) of this emission could be detected and
used to infer the state of the qubit. Repeated application of the transform
would produce a stream of emissions (i.e. a fluorescence ), thus increasing
the detection efficiency. Note that the existence of such a dissipative,
irreversible transition may be essential in order to implement quantum error
correction efficiently. Furthermore, our chosen representation of the qubit
basis states and the CU means that dissipitive transitions can be used to
prevent these objects from delocalizing \cite{longPaper}.

So far we have assumed that there is only one control unit (CU)\ in the
computer. Consequently we cannot apply a gate at several points
simultaneously. We could load an initial state containing $P$ CUs
distributed along the array (e.g. 1 every 20 qubits) \cite{LloydScience},
although we would then be constrained to apply exactly $P$ identical gates
simultaneously at {\em every} step, and always with the same spatial
distribution. Completely general parallelism would allow us to apply a
different number of simultaneous gates at each step, and at varying points
along the array. How can this ideal be approached? We cannot directly
create/annihilate CUs at specific locations because we are constrained to
use global updates. Figure 3 depicts an alternative solution which is
appropriate for any CA-like device. We increase the spacing between qubits
considerably, and in {\em each} space we put a CU and a set of classical
bits (using the same encoding employed for the qubits). Some of these
classical bits encode a label, for example we might label each space
uniquely using a binary number, and the others form an auxiliary `work-pad'.
Together the CU and the classical bits effectively constitute an entire
computer in the space between each qubit; we will refer to these as
`sub-computers'. Now suppose we are `running' a parallelized quantum
algorithm, and the next step calls for a specific 1- or 2-qubit gate $G$ to
be applied simultaneously at $P$ points along our array of $N$ qubits. This
operation would require $P$ CUs, located at {\em just} those points. However
we initially have one CU in {\em each} of the $N$ `sub-computers'. Therefore
we send an update sequence which causes a computation \cite{haveCCN}
simultaneously within each `sub-computer': the label bits are the input and
the output is a binary variable represented by some transformation
applied/not applied to the CU. This transformation disables the CU: when we
subsequently apply updates to move the CUs away from their sub-computers to
perform the gate operation $G$ on neighboring qubits, this will only occur
where the CUs are untransformed. One such transformation is shown in Fig.
3(b). Having thus implemented the step required by our parallel quantum
algorithm, we can now reverse the computation previously applied to the
`sub-computers' in order to return them to their initial state.

There are costs and constraints associated with using this procedure. The
size of the array must be increased by a factor $f$ due to the inclusion of
the `sub-computers'; unique labeling would imply $f$ of order $\ln (N)$. The
time $\tau $ associated with the `sub-computer' computation must be less
than $O(N)$ otherwise it would have been quicker to perform the parallel
gates in series. One could not enable/disable a completely arbitrary sub-set
of the $N$ CUs under this time constraint \cite{acknowledgeWim}, so our
procedure does not efficiently implement a completely general arrangement of
gates. However, there are a great many useful distributions of CUs which do
correspond to sufficiently fast $\tau $. The most obvious examples include:
all CUs, a given CU, one in every $2^{p}$ CUs, all CUs in some interval. For
these and many other patterns, $\tau $\ is merely of order $\ln (N)$. An
obvious variation is to place a `sub-computer' only every ten qubits, say.
Another is to `nest' the procedure to provide parallel computation within
the `sub-computers' at a cost of $\ln (\ln (N))$. The process performed by
the `sub-computer' could be generalized to apply a range of transformations
to the CU, corresponding to different subsequent gates operations on the
qubits. Most interestingly one could generalize the classical bits in each
`sub-computer' to qubits, so that the computation determining which CUs are
disabled becomes a quantum process producing CUs in a superposition of the
enabled/disabled state. It is unclear whether this could have significant
advantages for algorithmic efficiency.

In all the procedures described above the cells are only sensitive to their
immediate neighbors. In isolation a cell would have a certain energy gap
between its two states; in the array environment this is split into distinct
levels corresponding to the values of the field variable. However in a real
system the cells would also be influenced by the states of non-neighboring
cells, with the result that each level would be split into a multiplet of
many levels. In order to drive a transition in reasonable time it would be
desirable to address a multiplet collectively \cite{NMRpulse}; this could
only be achieved if the multiplets are non-overlapping. This condition
translates to a constraint on how short-range the physical cell-cell
interaction must be. It is easy to show that any one-dimensional system with
a symmetric interaction (right and left neighbors indistinguishable) has
multiplets that are well separated if the interaction is $r^{-3}$ or
shorter. Dipole elements such as nuclear or electron spins constitute one
class of examples. Note that when the multiplets are well separated, it is
possible to tolerate a degree of physical variation between cells which are
nominally of the same type, since the effect of such variation is merely to
broaden the multiplet correspondingly. Similarly, modest variations in the
inter-cell spacing and coupling strength could be tolerated. Note also that
if the interaction is not diagonal in the basis of the cell's states, the
scheme still functions provided that the difference between the fundamental
frequencies of the $A$\ and the $B$\ cells is large compared to the
magnitude of any off-diagonal terms \cite{sethLong}.

The present scheme is suited to systems where it is experimentally difficult
or impossible to target specific units for manipulation. One example
involves the nuclear magnetic resonance (NMR) approach \cite{NMR} which has
been successfully used to realize 3-qubit computers. Here the computer is a
molecule possessing a number of spin-non-zero nuclei, the states of which
are used to represent the qubits. Probably the most fundamental obstacle
preventing experimentalists from extending the number of qubits, $N$, is the
difficultly of distinguishing $N$ unique sets of energy levels. This
obstacle is removed by our model, which could be realized by a linear
molecule with $A$ and $B$ sites alternating along its length: we need not
distinguish between any two sites of a given type, hence we have only two
fixed sets of energy levels {\em regardless} of $N$. For a second example,
consider the solid state realization recently proposed by Kane \cite{Kane}.
Here qubits are again realized by the states of nuclear spins, but these
belong to donor impurity atoms embedded in Si. In order to gain control over
specific qubits, Kane introduces a set of electrostatic gates located near
the donors, with two gates being required for each donor. These electrodes
represent both a principal source of decoherence in the system, and a major
difficulty for experimental realization. By switching to the model presented
here, where there is no need to address qubits individually, the essential
role of the electrodes is removed. If, as seems entirely plausible \cite
{longPaper}, their remaining functions can be obviated by design
modifications, then it will be possible to dispense with \ them entirely.

To conclude, we have exhibited a model of quantum computation which requires
only global manipulations and yet has very modest physical requirements. We
have shown that it is possible to efficiently perform non-trivial parallel
operations on such a model. The model operates with interaction ranges as
great as $r^{-3}$, and is thus applicable to a wide range of QC
implementations where it may significantly reduce the obstacles to
experimental realization.

The author would like to thank Seth Lloyd, Mike Mosca and Wim Van Dam for
useful discussions. This work was supported by an EPSRC\ fellowship.

\bigskip

\newpage

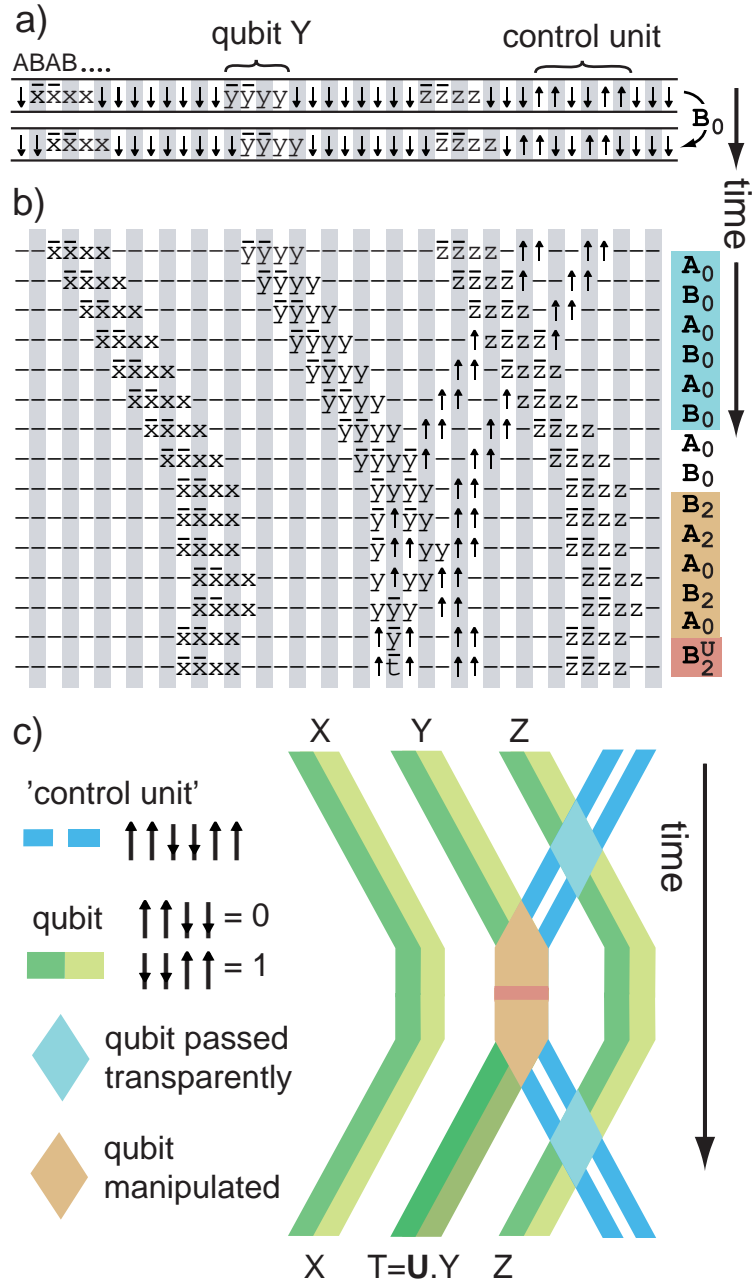
\noindent {\bf Fig.1 }A section of the array containing the control unit
(CU) and three qubits, $X$,$Y$\ \& $Z$, each encoded over four cells. All
other cells are in state `$\downarrow $'.{\bf \ }White cells are of type $A$%
, shaded cells of type $B$. (a) The effect of the update $B_{0}$: the CU
moves one cell to the left, all the qubits move one cell to the right, yet
the form of the qubits and the CU are preserved. (b) \& (c) A general
`one-qubit' gate. For clarity the states `$\downarrow $' are written as `$-$%
'. In response to the updates $A_{0}$,$B_{0}$,$A_{0}$,$B_{0}...$ the CU
passes through qubit $X$, leaving it {\em unchanged}, and continues until
mid-way through passing qubit $Y$. Then additional updates are applied: the
effect of the last is to apply a unitary transform ${\bf U}$ only to the
cell representing the $Y$ qubit, yielding qubit $T={\bf U}.Y$. As indicated
in (c), re-applying the updates in reverse order then moves the CU moves
away from $T$.

\bigskip

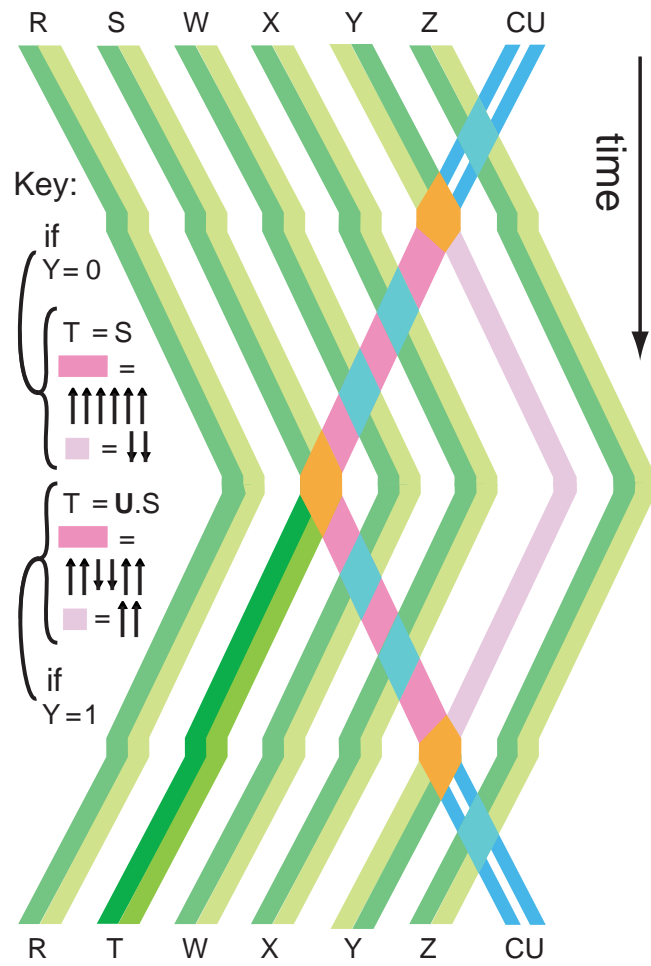
\noindent {\bf Fig.2 }Schematic of the `two-qubit' control-${\bf U}$
process. The target qubit is $S$, and the control is $Y$. The CU moves
transparently past the $Z$ qubit, and continues until mid-way through
passing $Y$. To this point is the process is identical to Fig 1(c), however
now the CU itself is subject to a transformation: it is altered from $%
\uparrow \uparrow \downarrow \downarrow \uparrow \uparrow $ to $\uparrow
\uparrow \uparrow \uparrow \uparrow \uparrow $ if, and only if, $Y=0$. Both
forms of the CU will pass transparently through the intervening qubits $W$,$%
X $ in answer to the same update sequence. When qubit $S$ is reached a new
sequence is applied, the last of which subjects $S$ to a transform ${\bf U}$%
\ if, and only if, the CU arrived in its unaltered form. Finally we re-apply
all the updates preceding the last\ {\em in reverse order} so as to return
the CU\ to its initial state. An explicit depiction of the process is
available from Ref. \cite{webFigs}.

\bigskip

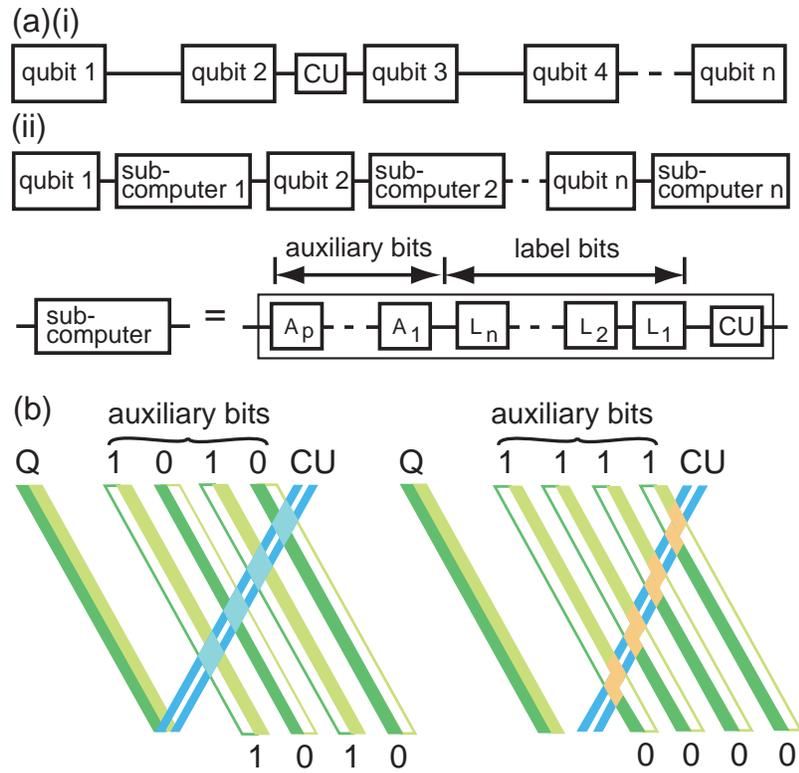
\noindent {\bf Fig.3 }(a) Generalization from the simple serial model (i) to
the parallel model (ii) employing `sub-computers'. (b) One means of
disabling the CU simply by delaying it. The CU is delayed (or not delayed)
depending only on the states of four auxiliary bits, which have been set by
the proceeding `sub-computer' computation. The sequence of updates applied
is the {\em same} for both cases. The delayed CU is in an `empty' region of
the array when the non-delayed CU has reached its target qubit, here denoted 
$Q$. An explicit depiction of the process is available from Ref. \cite
{webFigs}.

\bigskip \newpage

\noindent Caption for the Web Figures - Included on the web \cite{webFigs}.

\bigskip

\noindent {\bf NOT\ INTENDED\ FOR\ JOURNAL\ PUBLICATION\bigskip }

\noindent {\bf Web: \ FigureA.pdf.} The explicit description of the control-$%
{\bf U}$ process which is shown schematically in Figure 2. Note that the
update sequence is the same in both parts of the Figure, but only in the $%
Y=1 $ case does the last update, $B_{2}^{{\bf U}}$, have an effect. After
this update has been applied, the preceding updates would be re-applied in
reverse order to complete the process.

The two blue rows show the control unit having moved from the $X$ qubit to
the $W$ without changing its form. It follows that the CU\ could cross any
number of intervening qubits to reach its target.

Note: in this Figure we use only 4 spacer cells between each qubit; this is
the minimum that permits qubit gate operations. A consequence of this tight
packing is that one of the neighbors of the target qubit must be disturbed
during the operation (here $W$ is disturbed). However the final update, $%
B_{2}^{{\bf U}}$, affects only the target qubit and hence the disturbance of
the neighbor is undone when the preceding updates are re-applied in reverse
order.

\bigskip

\noindent {\bf Web: FigureB.pdf.} An explicit depiction of the delaying
transformation which is shown schematically in Figure 3(b). Note that the
transform is applied/not applied to the control unit depending only on the
values stored in the four auxiliary bits, i.e. the update sequence is the
same for both cases. The auxiliary bits will have been set by the proceeding
`sub-computer' calculation.

\bigskip

\noindent {\bf Web: FigureC.pdf.} The most compact implementation of the
important Control-Control-${\bf U}$ gate. The transformation ${\bf U}$ is
applied to the target qubit $W$ if, and only if, both the control qubits $X$
and $Y$ are in state 1.


\begin{references}
\bibitem{need2Types}  Using two alternating types guarantees that while a
cell is physically subjected to an update rule, its neighbors, on whom the
rule is defined, are static.
\bibitem{haveCCN}  This computation may be entirely classical. A suitable
universal classical gate can be obtained from the CCU of Ref. \cite{webFigs}
by choosing ${\bf U}$ to be a simple inversion (i.e. a NOT).
\end{references}
\end{document}